# On models to describe the volume in the context of establishing high-pressure Gibbs energy databases

Guillaume Deffrennes[a,*], Jean-Marc Joubert[b], Benoit Oudot[a,*]

[a] CEA, DAM, VALDUC, F-21120 Is-sur-Tille, France

[b] Univ. Paris Est Creteil, CNRS, ICMPE, UMR 7182, 2 rue Henri Dunant, 94320 Thiais, France

* Corresponding authors:

Dr. Guillaume Deffrennes

Present postal address: National Institute for Materials Science, 1-1 Namiki, Tsukuba, Ibaraki 305-0044, Japan

e-mail : guillaume.deffrennes@gmail.com

Dr. Benoit Oudot

Postal address: CEA, DAM, VALDUC, F-21120 Is-sur-Tille, France

e-mail : benoit.oudot@cea.fr

# Abstract

High-pressure Gibbs energy approaches are promising for establishing multi-component thermodynamic databases. However, so far, they have often been considered unsuccessful, because of their tendency to lead to unphysical extrapolations of the thermodynamic properties at elevated temperatures and pressures. Beyond this symptom, the root causes of the problem are rarely investigated. In this work, it is identified that these shortcomings are caused by (i) an inconsistent treatment of the SGTE method of extrapolation, and (ii) an insufficient knowledge of the models to describe the volume that are the key to extend CALPHAD databases toward high pressures. Because the first step toward solving the problem is to focus on the later issue, several models to describe the volume that are built upon the Grover empirical law are investigated, namely, the Lu-Grover, Joubert-Lu-Grover, Jacobs-Grover, and the Anand-Saxena-Grover approaches. In addition, an original description built upon the concept of thermal pressure, and a new scheme based on the equivalence between temperature and pressure when computing the volume are discussed. For each of these models, limitations and possibilities are identified from a practical standpoint.

# 1. Introduction

There is a strong interest in the establishment of multi-component thermodynamic databases valid up to high temperatures and pressures for applications not only in geophysics, but also in metallurgy [1–5]. To achieve this purpose, there are two main ways. On the one hand, the approaches that rely on the modeling of the Helmholtz energy as a function of temperature and volume [6–8] naturally relate to *ab initio* calculations. The models are built based on physical considerations, and simplifying assumptions. They allow to achieve reasonable descriptions up to extreme conditions of temperature and pressure for stoichiometric phases, but their application to the modeling of solution phases in multi-component systems have been limited [8]. They will not be further considered in the present study. On the other hand, the approaches that rely on the modeling of the Gibbs energy as a function of temperature and pressure naturally relate to experimental work. Therefore, the models tend to be empirical in nature, and phenomenological. They allow to achieve accurate descriptions of complex multi-component systems at atmospheric pressure within the framework of the CALPHAD method, but their extension toward high pressures met only limited success so far. That is because Gibbs energy approaches tend to lead to unphysical extrapolations at high pressure. For instance, a negative heat capacity was calculated in this range for Pt [6], Mo [9], W [10], or Al [11]. However, beyond this symptom, the root causes of the problem are rarely investigated. The Gibbs energy can, by definition, be expressed as the sum of two contributions as follows:

$$\Delta G(T,p) = \Delta G(T,p^0) + \int_{p^0}^{p} V(T,p')dp' \qquad (1.1)$$

with $\Delta G$ the Gibbs energy of a phase relative to a given reference state, $p$ the pressure and $p^0$ the atmospheric pressure, $T$ the temperature, and $V$ the molar volume. The composition dependence of the Gibbs energy and of the volume falls outside the scope of the present study.

It follows from Eq. (1.1) that the problematic extrapolations obtained at high pressures with Gibbs energy approaches can stem from either the description of the Gibbs energy at atmospheric pressure, or the one of the volume. It was discussed by Brosh *et al.* [11] that the origin of the problem was due to the so-called SGTE method of extrapolation [12]. Yet, it was recently highlighted that this was not the main cause of these shortcomings, and that the primary problem laid in the description of the volume and related properties [13].

In this work, the problem arising from the SGTE method of extrapolation is first discussed in Section 2. Then, a critical analysis of several models to describe the volume as a function of $T$ and $p$ is proposed. The focus is put on models built upon the empirical relationship discovered by Grover *et al.* [14] in the early 70s. To begin with, the model provided by Lu *et al.* [15] is investigated in Section 3. Then, a revised version of this model recently proposed by Joubert *et al.* [13] is discussed in Section 4. In Section 5, a theoretical analysis of the framework built by Jacobs *et al.* [16–19] is given. A revision of this Jacobs-Grover model is proposed in subsequent Section 6. Next, a description of the volume based on the Grover empirical law and on the use of thermal pressure is presented in Section 7. This model has the same general behavior as the one proposed by Anand *et al.* [20] based on the formulation from Saxena [21] of the Grover empirical law. Finally, a new scheme based on the equivalence between temperature and pressure is discussed in Section 8.

## 2. The problem arising from the SGTE method of extrapolation

The SGTE method of extrapolation [12], which is commonly used in CALPHAD databases, consists in setting the heat capacity of solids to an arbitrary constant value above their melting point to avoid their spurious re-stabilization at very high temperatures. This scheme is used because lattice instabilities are not accounted for within the CALPHAD framework, resulting in the Gibbs energy of solid phases being extrapolated into regions where it is not thermodynamically well-defined [22]. Yet, it seems challenging to both account for mechanical instabilities and model solution phases containing many elements using end-members.

From the Maxwell relation $(\partial C_p/\partial p)_T = -T(\partial^2 V/\partial T^2)_p$ and the definition of the volumetric thermal expansion $\alpha$, the following generally applicable equation is obtained:

$$\left(\frac{\partial C_p}{\partial p}\right)_T = -TV\left(\alpha^2 + \left(\frac{\partial \alpha}{\partial T}\right)_p\right) \qquad (2.1)$$

with $C_p$ the isobaric heat capacity.

In the models to describe the volume that are based on physical considerations, the temperature derivative of the thermal expansion coefficient always tend to increase with $T$ at high temperature. This is due to the rise of electronic, and eventually thermal vacancies contributions in this range [23]. Therefore, if these models are coupled with an atmospheric pressure CALPHAD database in which the heat capacity of solid phases is kept constant above their melting point, it follows from Eq. (2.1) that it will inevitably lead to negative $C_p$ at very high temperatures and pressures. This is why an "incompatibility" was observed by Brosh *et al.* [11] between the SGTE method of extrapolation and the Mie-Grüneisen equation of state. It is argued here that this problem is caused by an inconsistency more than an incompatibility. That is because the heat capacity, the thermal expansivity and the bulk modulus $K_T$ are closely related to each other, and their variations with temperature arise from the same underlying physics [24]. Therefore, if the SGTE method of extrapolation is applied to the heat capacity for practical purposes, it should also be extended to the description of $\alpha$ and $K_T$. In other words, if an arbitrary constant value is set for the heat capacity of solids above their melting point at atmospheric pressure, the same treatment should be applied to their thermal expansion coefficient and bulk modulus. Then, the description would be consistent, and it would be one way to solve the problem brought forward by Brosh *et al.* [11]. This approach would be practical and not physical, and, for a given model, its impact on the phase diagram and volumetric properties at high temperatures and pressures would have to be investigated. A more physical solution to solve this problem of inconsistency would be to remove the "SGTE constraint" from the databases, and then to rely on the high temperature extrapolations provided by the 2nd generation CALPHAD phenomenological descriptions, or preferably on the ones provided by the 3rd generation semi-empirical models. To avoid the spurious re-stabilization of solids at very high temperatures, the equal-entropy criterion proposed by Sundman *et al.* [25] could then be a remedial treatment.

However, to make any progress toward solving this problem of inconsistency, it is first necessary to have a suitable description of the volume, and a precise knowledge of how it behaves up to very high temperatures and pressures. Yet, such a knowledge is often lacking, because in the attempt to build high-pressure Gibbs energy databases so far, not a lot of attention has been given to the underlying description of the volume and its limitations.

# 3. Investigation of the Lu-Grover model

Grover *et al.* [14] found from static and dynamic compression data on a large variety of metals that, along isotherms, there was a nearly precise linear relationship between the molar volume and the logarithm of the isothermal bulk modulus. This empirical finding was observed up to volume changes of 40%, or up to pressure of two times the atmospheric bulk modulus. Subsequent studies suggest that this empirical law is not specific to metals, but also apply, at least to some extent, to oxides such as MgO and $Mg_2SiO_4$ [16,18], and ionic compounds such as NaCl [20].

Based on the Grover empirical law, an explicit formulation of the volume was proposed by Lu *et al.* [15] as follows:

$$V = -cEi^{-1}\left(Ei\left(-\frac{V^0}{c}\right) - \frac{1}{K_T{}^0}\exp\left(-\frac{V^0}{c}\right)(p-p^0)\right) \qquad (3.1)$$

with $c$ a material characteristic parameter that can be temperature dependent, $K_T$ the isothermal bulk modulus, and $Ei$ the exponential integral function that can be calculated numerically from tabulations. More details on how Eq. (3.1) is derived from the empirical finding from Grover *et al.* [14] are provided in Supplementary Note A. It is noted that, for the sake of clarity, a simple nomenclature is used for equations: the independent variables $T$ and $p$ do not appear explicitly, and for instance $V(T,p)$ is simply written $V$. Besides, the underscript "0" is referring to the reference temperature, and superscript "0" to the atmospheric pressure, and for instance $K_T(T,p^0)$ is written $K_T{}^0$.

It was demonstrated by Lu *et al.* [15] that an explicit expression of the contribution to the Gibbs energy from the volume can then be obtained as follows:

$$G - G^0 = cK_T{}^0 \left(\exp\left(\frac{V^0 - V}{c}\right) - 1\right) \qquad (3.2)$$

The model proposed by Lu *et al.* [15] was applied in this work to the β-Sn phase. The atmospheric pressure description of the volume and of the isothermal bulk modulus was taken from [24]. A comparison between this description [24] and the experimental data [26–50] is presented in Supplementary Notes A and B. It is noted that this description is based on a novel 3rd generation CALPHAD model which differs from the phenomenological polynomial functions of $T$ used so far in Gibbs energy approaches. However, numerically speaking, very similar descriptions could be obtained above room temperature using either type of model. Therefore, it is emphasized here that the results obtained throughout this study are not exclusive to the 3rd generation CALPHAD framework. A value for the $c$ parameter of 2.945x10$^{-6}$ was obtained from the fit of the molar volume [34,35,51,52] and bulk modulus [53–56] data available along the room temperature isotherm. The agreement that was reached is presented in Supplementary Note C. The Gibbs energy description at atmospheric pressure was taken from the 3rd generation CALPHAD modeling proposed by Khvan *et al.* [57], which is supported by abundant and consistent data [57–65], as shown in Supplementary Note B.

On this basis, the isobaric heat capacity of β-Sn was calculated at various pressures using the Thermo-Calc software [66], and the results are presented in Fig. 1. A reasonable trend is obtained up to roughly 10 GPa, after which abnormal results are obtained. Indeed, the heat capacity becomes negative at higher pressure, which will lead to a negative entropy. It appears clearly from Fig. 1 that the so-called SGTE method of extrapolation is not at the origin of these abnormal results, as negative values for the heat capacity are obtained below 505 K, which is the atmospheric pressure melting point of the phase. Because the atmospheric pressure description proposed by Khvan *et al.* [57] is based on solid grounds (Fig. S1), it follows from Eq. (1.1) that the contribution to the Gibbs energy from the volume is the source of the problem.

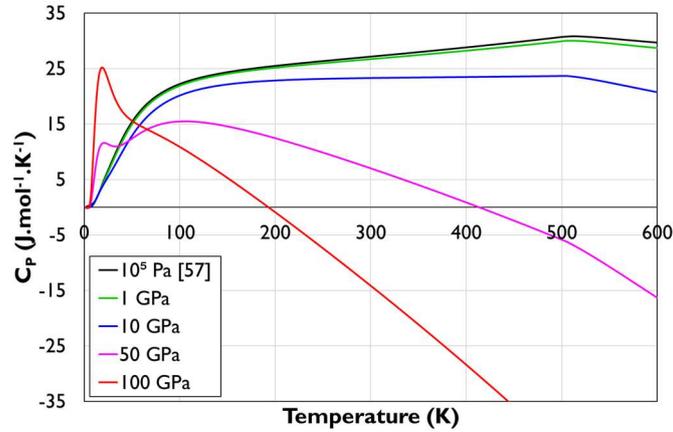

Fig. 1 – Isobaric heat capacity of β-Sn calculated using the Lu-Grover model

It follows from Eq. (2.1) that the problematic heat capacity extrapolations obtained at high pressure within the Lu-Grover framework must come from an unphysical description of the thermal expansion coefficient in this range. A hypothesis often made for solids is that the product of the thermal expansion by the bulk modulus is temperature independent [67]. It was claimed by Kumar *et al.* [68,69] that this product was also pressure independent, although no justification was provided by the authors to support this assumption. It was noted by Anderson *et al.* [70] that this hypothesis was also considered by Birch [71] for oxides and silicates based on experiments on alkali metals to up to 3 GPa. Therefore, this simplifying assumption is not well-established, and further investigations, possibly from *ab initio* calculations, would be of interest. Yet, this constraint may be useful to avoid obtaining an abnormal description of the thermal expansion at high pressure. Indeed, as the bulk modulus increases with pressure, the thermal expansion coefficient would naturally tend toward 0, which is the expected trend. Besides, the product $\alpha K_T$ is also closely related to the thermal pressure, a property from which equations of state can be built upon, as it will be discussed in Section 7. For both these reasons, the variations of this product with pressure will be investigated for each model considered in this work, starting with the Lu-Grover framework. For this model, the following equation can be obtained, as demonstrated in Supplementary Note A:

$$\left(\frac{\partial \alpha K_T}{\partial p}\right)_T = \frac{V^0 \alpha^0}{c} + \frac{1}{K_T^0}\left(\frac{\partial K_T^0}{\partial T}\right)_p - \frac{1}{c}\ln\left(\frac{K_T}{K_T^0}\right)\left(\frac{\partial c}{\partial T}\right)_p \tag{3.3}$$

In the present case of β-Sn, variations along different isobars of the product $\alpha K_T$, of $\alpha$ and of $K_T$ are presented in Fig. 2. As the positive $c$ parameter is here a constant, the last term of Eq. (3.3) is removed. As a result, it can be deduced from Eq. (3.3) that the product $\alpha K_T$ varies linearly with pressure along isotherms. Plus, Eq. (3.3) now simplifies into two contributions: a positive one linked to the description of the volume at atmospheric pressure, and a negative one linked to the temperature derivative of the bulk modulus at atmospheric pressure. It can be seen from Fig. 2(a) that at 437 K, the product $\alpha K_T$ is pressure independent, which means both remaining terms of Eq. (3.3) cancel each other out. At lower temperatures, $\alpha K_T$ decreases linearly with increasing pressure, and becomes negative on an increasingly wide temperature range. Thus, so does the thermal expansion, as seen in Fig. 2(b). It follows from Eq. (2.1) that this is what causes the spurious low temperature bumps observed above 10 GPa on the heat capacity of β-Sn in Fig. 1. At higher temperatures, the product $\alpha K_T$ increases linearly with increasing pressure. As a result, the higher the pressure, the higher the increase of thermal expansion with temperature. This result is also unphysical, and following Eq. (2.1), it leads to the negative heat capacity observed in Fig. 1.

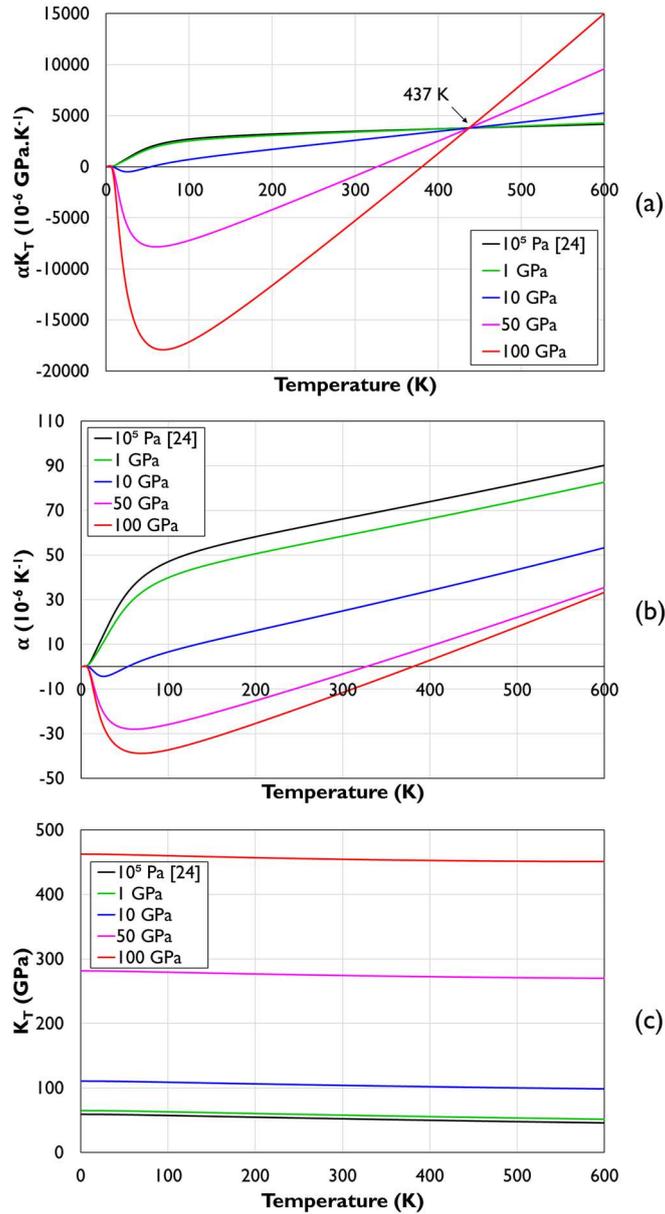

Fig. 2 – (a) Product $\alpha K_T$, (b) thermal expansion coefficient, and (c) bulk modulus of β-Sn calculated along different isobars using the Lu-Grover model

At atmospheric pressure, the modeling of β-Sn is well-constrained by the data (Fig. S1-S2). Using the Lu-Grover model, the data on the volume available up to 15 GPa are reproduced closely (Fig. S2), but anomalies in the thermal expansion appear from 10 GPa (Fig. 2). In a recent study, the Lu-Grover model was applied to the Ti unary, and data on phase equilibria and volumetric properties were satisfactorily reproduced up to at least 20 GPa [5]. Using the authors' description, negative values for the thermal expansion coefficient of hexagonal close-packed Ti are obtained from 100 GPa. The experimental data on the volumetric properties of MgO were satisfactorily reproduced by Lu *et al.* [15], including thermal expansion data available up to 2000 K and 200 GPa. Yet, the description of the thermal expansion of the phase causes the $C_p = f(T)$ curve to take an inverted U-shape from 50 GPa. It is concluded that the Lu-Grover model is only applicable at relatively low pressures that should not exceed one fourth of the standard bulk modulus of the phases. Nonetheless, despite this limitation, this model can help in understanding microstructure evolution in metals subjected to severe plastic deformation [5].

# 4. Investigation of the Joubert-Lu-Grover model

In an analysis complementary to the one presented in Section 3, Joubert *et al.* [13] recently investigated on the origin of the unphysical extrapolations obtained at high pressure within the Lu-Grover framework. It was demonstrated that abnormal heat capacity extrapolations arise from the uncontrolled behavior of the thermal expansion at high pressure, and to a lesser extent from the cross correlations between thermal expansion and compressibility. A revision of the Lu-Grover model was proposed so that a reasonable description of the volume would be obtained at high pressure. This revised approach was applied to the description of the Os-Pt binary system [72].

In the Joubert-Lu-Gover model, the description of the volume is still provided by Eq. (3.1), following the work of Lu *et al.* [15]. However, pressure dependent cut-off parameters are introduced in the expressions of $\alpha^0$ and of $K_T^0$ as follows:

$$\alpha^0 = A_0 \exp\left(-\frac{p}{p_{CUT}'}\right) + \sum_i A_i T^i \exp\left(-\frac{p}{p_{CUT}}\right) \tag{4.1}$$

$$K_T^0 = \frac{1}{B_0 + \sum_i B_i T^i \exp\left(-\frac{p}{p_{CUT}}\right)} \tag{4.2}$$

with $p_{CUT}$ and $p_{CUT}'$ the added cut-off parameters, $A_i$ and $B_i$ the fitting parameters of the polynomial functions used for 2$^{nd}$ generation CALPHAD descriptions of $\alpha^0$ and $K_T^0$, and $i$ an integer. It is noted that in the original model from Lu *et al.* [15], $\alpha^0$ and $K_T^0$ are the atmospheric pressure thermal expansion coefficient and bulk modulus, but within the Joubert-Lu-Gover framework, at high pressures this is not true anymore, and they should then be considered more as model parameters.

It appears clearly from plotting $\exp(-x/x_{CUT})$ as a function of $x/x_{CUT}$ that significant variations with $x$ are only obtained in the $10^{-3}\,x_{CUT} < x < 10\,x_{CUT}$ range. Consequently, the pressure cut-off terms added in Eq. (4.1) and (4.2) such that $p_{CUT} < p_{CUT}'$ only lead to significant variations with pressure in the $10^{-3}\,p_{CUT} < P < 10\,p_{CUT}'$ range. Hence, outside this range $\alpha^0$, $V^0$, as well as $K_T^{\,0}$ are pressure independent, and all the equations obtained within the Lu-Grover framework are still valid. When the pressure is lower than $10^{-3}\,p_{CUT}$, both cut-off terms are basically equal to 1, and the very same behavior as in the Lu-Grover model is maintained. When the pressure is greater than $10\,p_{CUT}'$ however, both cut-off terms are basically equal to 0, and provided that the $c$ parameter is a constant, it can be shown that the thermal expansion becomes null, and the bulk modulus temperature independent. It is noteworthy that if the $c$ parameter is modeled as a polynomial function of $T$ as in the original work of Lu *et al.* [15], it can be deduced from Eq. (3.3) that, to keep the product $\alpha K_T$ well-constrained, a cut-off parameter should also be applied to it as follows:

$$c = C_0 + \sum_i C_i T^i \exp\left(-\frac{p}{p_{CUT}}\right) \tag{4.3}$$

with $C_i$ the fitting parameters of the polynomial function used to describe the $c$ parameter.

A consequence of the modifications presented in Eq. (4.1), (4.2) and (4.3) is that, in the Joubert-Lu-Grover model, $\alpha^0$, $K_T^{\,0}$ and possibly also $c$ are pressure dependent in the $10^{-3}\,p_{CUT} < p < 10\,p_{CUT}'$ range. As a result, the explicit expression of the contribution to the Gibbs energy from the volume presented in Eq. (3.2) does not hold anymore. Therefore, in order to compute the thermodynamic properties at high pressure, it is required to perform a numerical integration of $V$ over $p$. Yet, it may still be interesting to use Eq. (3.2). Besides from allowing to gain in computational efficiency, it can be shown that within the Joubert-Lu-Grover framework, all the parameters from Eq. (3.2) are temperature independent at high pressure. Therefore, if Eq. (3.2) is used, there will be no contribution from the volume to the entropy in the high-pressure range, meaning that the heat capacity will start by decreasing with pressure, but will eventually increase back up to its atmospheric pressure value. This trend is unphysical, and the Gibbs energy function calculated that way is inexact. Yet, to follow this approach has the benefit of ensuring direct compatibility with the SGTE method of extrapolation. Therefore, it could be a be a practical solution to the problem of inconsistency that was discussed in Section 2, leading at high pressure and temperature to an approximate but reasonable thermodynamic description, and maybe to a correct phase diagram as suggested by recent studies [72,73].

A practical investigation of the Joubert-Lu-Grover model was conducted taking the case of β-Sn. As in Section 3, the description at atmospheric pressure of the thermal expansion and the bulk modulus were accepted from [24]. Because it is based on a 3rd generation CALPHAD model, the Joubert-Lu-Grover model that was proposed in the framework of 2nd generation polynomial formulations had to be adapted as described in detail in Supplementary Note D. In order to adjust the $p_{CUT}$ and $p_{CUT}'$ parameters, information on the volume of the phase in the $T$ - $p$ space are ideally required. Yet, high-pressure measurements for the β-Sn phase are limited along the room temperature isotherm. Therefore, an arbitrary value of $10^{10}$ Pa was selected for $p_{CUT}'$. Then, values for $p_{CUT}$ and $c$ of respectively $2 \times 10^9$ Pa and $3.456 \times 10^{-6}$ were obtained from fitting the bulk modulus [53–56] and molar volume [34,35,51,52] data available along the room temperature isotherm. The agreement that was reached is presented in Supplementary Note E. It is noted that different values for the parameters listed above could lead to equally satisfying results, and additional data at high pressure, possibly from DFT calculations, would be needed to constrain further the model. Nonetheless, the available experimental data are enough for the present purpose, which is to highlight the general features of different models. Finally, the atmospheric pressure description of the Gibbs energy was taken from [57].

From this description, various calculations were conducted along the 298.15 K isotherm using the Thermo-Calc software [66] and worksheets, and the results are presented in Fig. 3. It is shown in Fig. 3(a) that the product $\alpha K_T$ reaches a constant value of 0 at high pressure within the Joubert-Lu-Grover framework, whereas it becomes negative for the description based on the Lu-Grover model. Because the bulk modulus is similar for both modeling as shown in Fig. 3(b), this change in the product $\alpha K_T$ mainly impacts the thermal expansion, that is presented for both cases in Fig. 3(c). Using the Joubert-Lu-Grover model, the thermal expansion does not exhibit an abnormal behavior at high pressure anymore, but does tend to 0, which is the expected and reasonable trend.

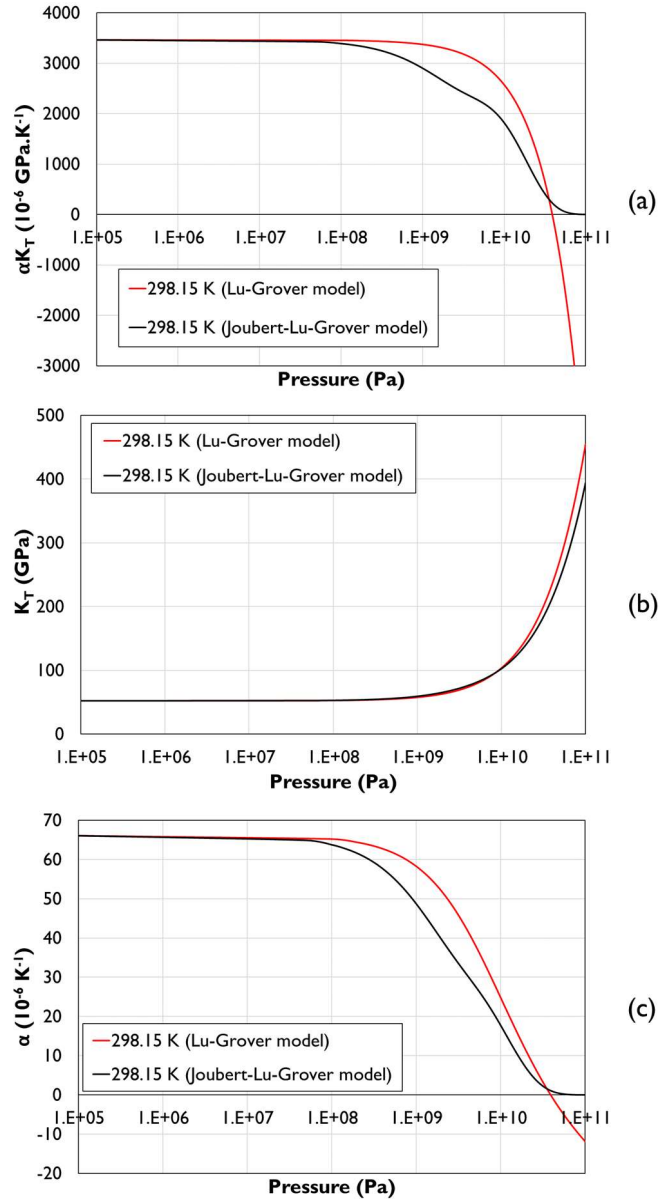

Fig. 3 – (a) Product $\alpha K_T$, (b) bulk modulus, (c) thermal expansion coefficient of β-Sn calculated along the 298.15 K isotherm using the Joubert-Lu-Grover model and the Lu-Grover model.

The Joubert-Lu-Grover model has been applied successfully to describe up to high temperatures and pressures the volumetric properties of face-centered cubic Pt [13], hexagonal close-packed Os [72], and body-centered tetragonal and cubic Sn [73]. Yet, after calculating the Gibbs energy by numerical integration of $V$ over $p$, it appears from Fig. 4(a) that the heat capacity is overestimated at high temperatures and pressures, considering that it should converge to the Dulong-Petit limit of 3R in this range. This overestimation is due to the steep decline of the temperature derivative of the thermal expansion coefficient above the cut-off term $\exp(-p/p_{cut})$. A possible solution to improve the description would be to use a different mathematical expression for this term. Another possibility is, as discussed above, to use Eq. (3.2) to compute the thermodynamic functions instead of performing a numerical integration of the volume. Doing so, it appears from Fig. 4(b) that, as expected, at high pressures the $C_p$ increases back toward its atmospheric pressure value. While this trend is abnormal, this approach may still be a practical solution for extending CALPHAD databases toward high pressures if the "SGTE constraint" (Section 2) is set to the Dulong-Petit limit. Indeed, this approach was applied to the Sn unary, and the available data on phase equilibria, shock compression and volumetric properties were closely reproduced up to pressures of almost 3 times the standard bulk modulus of the element, and temperatures 5 times higher than its atmospheric pressure melting point [73].

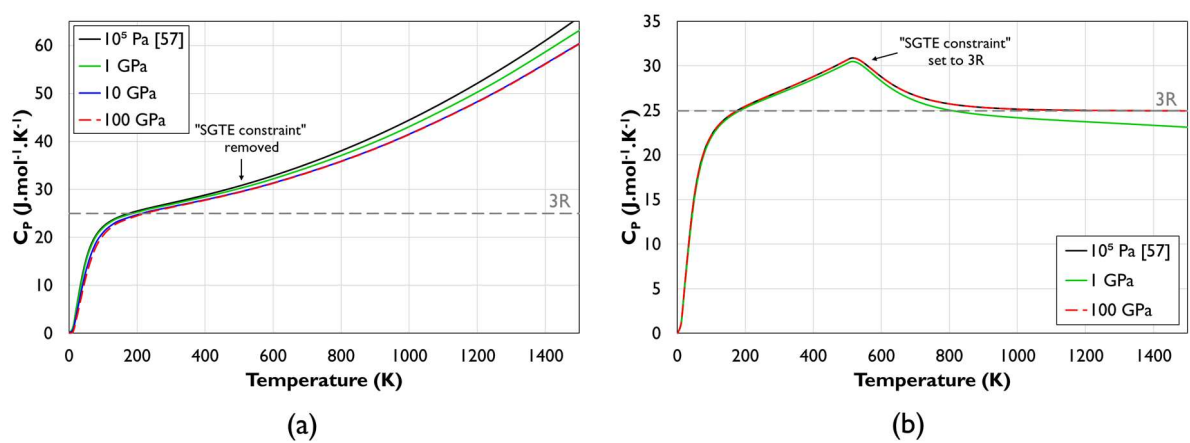

(a)          (b)

Fig. 4 – Heat capacity of β-Sn calculated in two different ways using the Joubert-Lu-Grover model. (a) Exact determination from the Gibbs energy that was calculated by numerical integration of the volume, and (b) approximate determination from Eq. (3.2). In (a), the description of the heat capacity at atmospheric pressure was left unconstrained above the melting point of the phase of 505 K, whereas in (b) the "SGTE constraint" was set to 3R.

# 5. Investigation of the Jacobs-Grover model

The empirical relation discovered by Grover *et al.* [14] was originally an isotherm. Yet, Jacobs *et al.* [16–19] assumed that this relationship was also valid along isobars. The following relationship was then proposed by the authors in their former work on MgO [16,17]:

$$V = V_0^0 - c \ln\left(\frac{K_T}{K_{T_0}^0}\right) \tag{5.1}$$

It is highlighted that the parameters of Eq. (5.1) are constants, as they are independent not only of $p$ as in the Lu-Grover model, but also of $T$.

In their later work on $Mg_2SiO_4$ [18] and $Fe_2SiO_4$ [19], Jacobs *et al.* introduced an additional parameter in Eq. (5.1), leading to:

$$V = V_0^0 + a(T - T_0) - c \ln\left(\frac{K_T}{K_{T_0}^0}\right) \tag{5.2}$$

with $a$ the additional material-dependent constant, which role will be explained shortly.

By re-arranging Eq. (5.2), following the same approach detailed in Supplementary Note A for the Lu-Grover model, the following differential equation is obtained:

$$\frac{\exp\left(-\frac{V_0^0 + a(T - T_0)}{c}\right)}{K_T^0} = -\frac{\exp\left(-\frac{V}{c}\right)}{V}\left(\frac{\partial V}{\partial p}\right)_T \tag{5.3}$$

Unlike Lu *et al.* [15], Jacobs *et al.* [19] did not use the exponential integral function to solve Eq. (5.3), but a power series, leading to:

$$p = p^0 - K_{T_0}^0 \exp\left(\frac{V_0^0 + a(T - T_0)}{c}\right)\left(\ln\left(\frac{V}{V^0}\right) + \sum_{j=1}^{\infty}\left(-\frac{c^{-j}(V^j - (V^0)^j)^j}{j \times j!}\right)\right) \tag{5.4}$$

An explicit expression of the volume cannot be obtained from Eq. (5.4), and roughly 20 to 30 terms are needed in the power series to obtain accurate descriptions. Therefore, it makes the Jacobs-Grover approach less computationally efficient compared to the models investigated so far. Besides from this practical consideration, let us now investigate the implications of the fact that, in the model of Jacobs *et al.* [16–19], the relation from Grover *et al.* [14] was considered to hold along both isotherms and isobars, i.e., the reference volume and bulk modulus are constants in the initial Eq. (5.1). As detailed in Supplementary Note F, the following equations can be obtained:

$$\left(\frac{\partial \alpha K_T}{\partial p}\right)_T = \frac{a}{c} \tag{5.5}$$

$$\left(\frac{\partial K_T}{\partial T}\right)_V = \frac{a}{c} K_T \tag{5.6}$$

It appears from Eq. (5.6) that without using the $a$ parameter, the bulk modulus would be temperature independent along isochores. As a result, it follows from Eq. (5.5) that the product $\alpha K_T$ would be pressure independent along isotherms, as discussed by Jacobs *et al.* [74] in their reply to the comments from Raju *et al.* [75]. The $a$ parameter is therefore needed to account for the deviations from this ideal behavior. In practice, it can also be shown that without using the $a$ parameter, the $c$ parameter solely accounts for the variations of the bulk modulus along both isobars and isotherms. A satisfying fit of the available data may therefore not be obtained with this single degree of freedom, as it was notably highlighted by Jacobs *et al.* [18] in the case of $Mg_2SiO_4$. Nonetheless, if the $a$ parameter is negative, such as for $Mg_2SiO_4$ [18], the product $\alpha K_T$ will exhibit a monotonic decrease with increasing pressure. At high pressure, this will inevitably result in a negative thermal expansion coefficient. This feature may lead to similar problems as the ones encountered with the Lu-Grover model (Section 3).

# 6. A revised Jacobs-Grover model

In section 5, it was highlighted that, although the Jacobs-Grover model leads to a more predictable behavior at high pressure than the Lu-Grover approach, it is not free from abnormal behavior nonetheless. Besides, it is also less computationally efficient. In the following, a revised Jacobs-Gover model is proposed, aiming at solving both issues.

First of all, in order to avoid obtaining a monotonic decrease of the product $\alpha K_T$ with increasing pressure, a pressure dependent cut-off term, similar to the ones introduced in the Joubert-Lu-Grover model, is added on the $a$ parameter of Eq. (5.2) as follows:

$$a = a^0 \exp\left(-\frac{p}{p_{CUT}}\right) \tag{6.1}$$

with $a_0$ the initial parameter optimized using available bulk modulus data, and $p_{CUT}$ the pressure cut-off from which the $a$ parameter will rapidly drop to 0. Following the same approach as detailed in Supplementary Note A and E, it can be shown that:

$$\left(\frac{\partial \alpha K_T}{\partial p}\right)_T = \frac{a}{c} + \frac{\alpha(T-T_0)}{c}\left(\frac{\partial a}{\partial p}\right)_T \tag{6.2}$$

As discussed in Section 4, the cut-off term added in Eq. (6.1) basically acts as a switch which goes from 1 for pressures below $10^{-3}\, p_{CUT}$ to 0 for the ones above $10\, p_{CUT}$. Therefore, when $p < 10^{-3}\, p_{CUT}$, the $a$ parameter is equal to the initial value $a_0$ and is pressure independent, so the same behavior as the original Jacobs-Grover model is obtained. When $p > 10\, p_{CUT}$ however, the $a$ parameter becomes equal to 0, and so does its pressure derivative. As a result, it can be seen from Eq. (6.2) that the product $\alpha K_T$ will become pressure independent as well. Therefore, the product $\alpha K_T$ will not exhibit a monotonic decrease at high pressure anymore, and the modification proposed in Eq. (6.1) thus fixes the anomaly of the Jacobs-Grover model highlighted in Section 5.

Then, instead of using a power series to solve differential equation (5.3), the exponential integral function can be used. As in the Lu-Grover approach, an explicit expression of the volume can then be obtained. By injecting pressure dependent Eq. (6.1) into Eq. (5.3), it turns out that the exponential integral function now has to be used to solve both sides of the differential equation. By re-arranging the obtained result, the following description of the volume is finally obtained:

$$V = -c\, Ei^{-1}\left( Ei\left(-\frac{V^0}{c}\right) \right.$$

$$-\frac{1}{K_{T_0}^0}\exp\left(-\frac{V_0^0}{c}\right) p_{CUT}\left( Ei\left(-\frac{a^0(T-T_0)}{c}\exp\left(-\frac{p^0}{p_{CUT}}\right)\right)\right.$$

$$\left.\left.- Ei\left(-\frac{a^0(T-T_0)}{c}\exp\left(-\frac{p}{p_{CUT}}\right)\right)\right)\right) \quad (6.3)$$

Despite appearances, Eq. (6.3) is no more complicated than Eq. (3.1) from the Lu-Grover model, and it can be solved in a worksheet after tabulating the exponential integral function. In practice, a numerical problem is however obtained at pressures significantly higher than $p_{CUT}$, because $\exp(-p/p_{CUT})$ then quickly leads to unreasonably low arguments for the most right hand side $Ei$ function. To overcome this practical issue, we used the fact that when $p > 20\, p_{CUT}$, the $a$ parameter defined in Eq. (6.1) can very reasonably be taken to be 0, as it is $10^9$ times smaller than its original value at $p^0$. Therefore, for pressures higher than $20\, p_{CUT}$, differential equation (5.3) can be solved considering both distinct intervals as follows:

$$V = -cEi^{-1}\left(Ei\left(-\frac{V^0}{c}\right)\right.$$

$$-\frac{1}{K_{T_0}^0}\exp\left(-\frac{V_0^0}{c}\right)p_{CUT}\left(Ei\left(-\frac{a^0(T-T_0)}{c}\exp\left(-\frac{p^0}{p_{CUT}}\right)\right)\right.$$

$$\left.\left.-Ei\left(-\frac{a^0(T-T_0)}{c}\exp(-20)\right)\right) + p - 20p_{CUT}\right) \quad (6.4)$$

Finally, for materials for which $a$ is equal to 0 at atmospheric pressure, such as MgO [16,17], the description of the volume is provided by the following expression, reminiscent of Eq. (3.1) from the Lu-Grover model:

$$V = -cEi^{-1}\left(Ei\left(-\frac{V^0}{c}\right) - \frac{1}{K_{T_0}^0}\exp\left(-\frac{V_0^0}{c}\right)(p-p^0)\right) \quad (6.5)$$

Following the same procedure that was presented by Lu *et al.* [15] to derive their Eq. (10), the effect of pressure on the Gibbs energy function is obtained from Eq. (5.3) as follows:

$$G - G^0 = cK_{T_0}^0 \exp\left(\frac{V_0^0 + a^0 \exp\left(-\frac{p}{p_{CUT}}\right)(T-T_0)}{c}\right)\left(\exp\left(-\frac{V}{c}\right) - \exp\left(-\frac{V^0}{c}\right)\right) \quad (6.6)$$

This revised Jacobs-Grover equation of state was applied to the modeling of the β-Sn phase. The description of $V^0$ and $G^0$ were taken from [24] and [57], respectively. Values for the $c$, $a$ and $p_{CUT}$ parameters of 3.5x10⁻⁶, -6x10⁻¹⁰ and 10⁹ Pa were respectively obtained from fitting the bulk modulus [53–56] and molar volume [34,35,51,52] data along the room temperature isotherm, as well as the bulk modulus data along the atmospheric pressure isobar [32,48–50] critically selected in [24]. A satisfying agreement between the model and the available experimental data was reached, as presented in Supplementary Note G.

From the obtained description, calculations were conducted along different isobars, and the results are presented in Fig. 5. To begin with, it is highlighted in Fig. 5(a) that, in contrast with the Lu-Grover model, the product $\alpha K_T$ does vary only slightly with increasing pressure, and becomes pressure independent above $p_{CUT}$ due to our modification. However, a very different trend is noted regarding the variations of this quantity with temperature, which are imposed by the Grover empirical law that was considered valid not only along isotherms, but also along isobars. Indeed, it appears from Fig. 5(a) that above 1000 K, the product $\alpha K_T$ starts decreasing strongly with temperature. Then, from 3500 K, it slowly diminishes down to 0 GPa.K$^{-1}$ at atmospheric pressure, or down to a slightly negative value at higher pressure due to the use of the $a$ parameter. The bulk modulus presented in Fig. 5(b) varies with temperature in a somewhat consistent manner, decreasing steeply before also becoming almost temperature independent above 3500 K. Next, it is shown in Fig. 5(c) that when the pressure increases even so slightly, there is a critical temperature from which the thermal expansion coefficient will suddenly drop from the accepted atmospheric pressure description. This critical temperature decreases with increasing pressure. Indeed, if the anomaly occurs from roughly 4700 K just above atmospheric pressure, at 100 GPa this behavior is obtained from 1000 K, which is roughly twice the atmospheric pressure melting point of the phase. It follows from Eq. (2.1) that this decrease in the thermal expansion coefficient will lead to a significant increase in the heat capacity at high pressure, as highlighted in Fig. 5(d), which is unphysical. This limitation is inherent to the Jacobs-Grover framework, and comes from the fact that the Grover relationship was considered to be valid along isobars, a hypothesis which does not seem to hold at high temperatures, especially when the pressure is also high. Using the description of MgO proposed by Jacobs and Oonk [16,17], it is found that the temperature derivative of the thermal expansion coefficient of the phase becomes negative from 4000 K at 10 GPa, and from 2500 K at 160 GPa. It is concluded that, at pressures close to the standard bulk modulus of the material, the Jacobs-Grover model is only applicable up to relatively low temperatures that should not exceed its atmospheric pressure melting point.

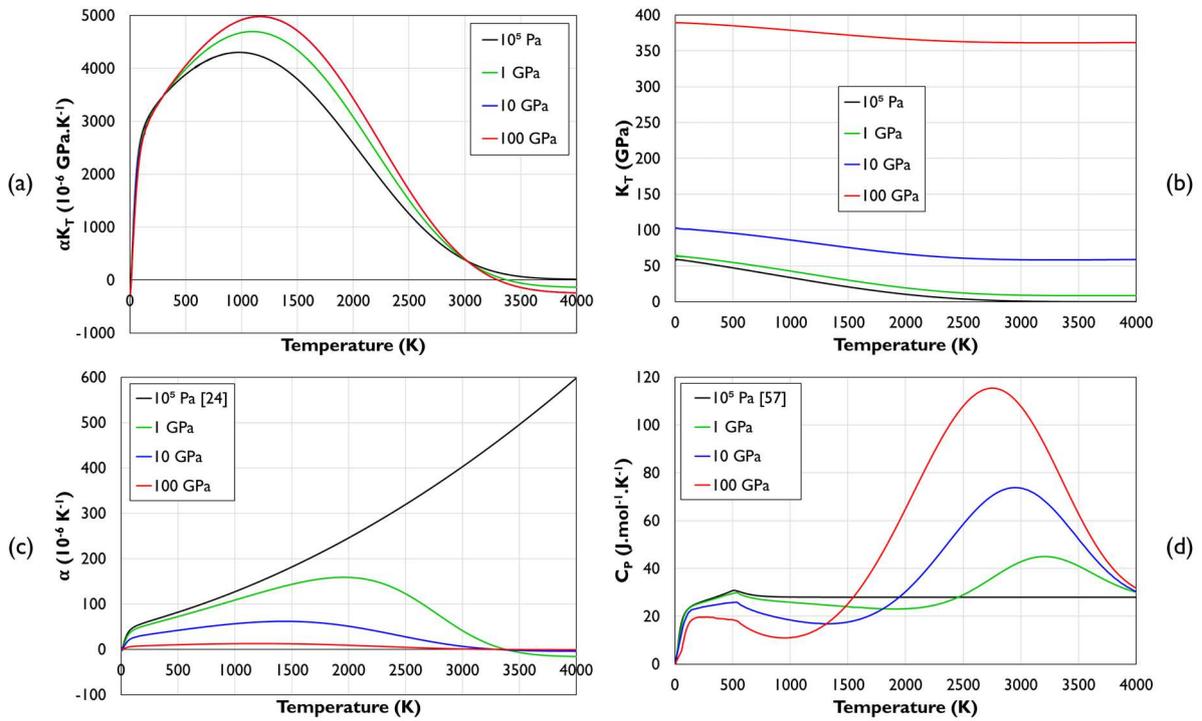

Fig. 5 – (a) Product $\alpha K_T$, (b) bulk modulus, (c) thermal expansion coefficient, and (d) heat capacity of β-Sn calculated along different isobars using the revised Jacobs-Grover model. The discontinuity in the temperature derivative of the heat capacity function in (d) at 505 K is caused by the SGTE method of extrapolation.

## 7. Investigation of a model based on the use of thermal pressure

In the models investigated so far, the effect of temperature on the volume were accounted for by injecting temperature dependent descriptions of $\alpha$, $V$ and $K_T$ into isothermal equations of state. In other words, the base principle was to first take position along the reference isobar by computing the volume and related properties at the temperature of interest, and then to extend the description toward high pressure from this point. An alternative to this approach is to make use of the concept of thermal pressure, which was extensively discussed by Anderson [67]. In equations of state in the form $p = f(T,V)$, it is common to express the pressure as the sum of two contributions. The first one, referred as the cold pressure, is computed as a function of the volume only using an isothermal equation of state. The second one is the thermal pressure, noted $p_{TH}$. It is obtained based on the thermodynamic identity $(\partial p/\partial T)_V = \alpha K_T$ from the following integral, which is made at constant volume:

$$p_{TH} = \int_{T_0}^{T} \alpha K_T dT \qquad (7.1)$$

For the present purpose, an equation of state in the form $V = f(T,p)$ is required. Then, it follows from the definition of thermal pressure that the volume at given high temperature and pressure conditions is equivalent to a volume at a lower pressure along the reference isotherm:

$$V(T,P) = V(T_0, p - p_{TH}) \qquad (7.2)$$

In practice, in order to solve Eq. (7.1) without using $V$ as an independent variable, hypotheses have to be made. For instance, Anand *et al.* [20] considered that the product $\alpha K_T$ was a constant, independent of both $T$ and $p$.

In this work, the temperature dependence of the product $\alpha K_T$ is modeled within the framework proposed in [24], which is built upon on a multi-frequency Einstein-Grüneisen model. The following equation is obtained:

$$\alpha K_T = \frac{3R}{V_0^0} \sum_i \gamma_{i_0} a_i \left( \left(\frac{\theta_i}{T}\right)^2 \frac{e^{\frac{\theta_i}{T}}}{\left(e^{\frac{\theta_i}{T}} - 1\right)^2} + AT + BT^2 \right) \tag{7.3}$$

with $R$ the gas constant, $\theta_i$, $a_i$ and $\gamma_{i_0}$ the Einstein temperature, a pre-factor, and the Grüneisen parameter associated with the i$^{th}$ Einstein mode of vibration, and $A$ and $B$ parameters to account for anharmonic and electronic contributions to the heat capacity.

Then, the product $\alpha K_T$ is considered to be pressure independent, and it follows from Eq. (7.1) that the thermal pressure can be obtained by integration of Eq. (7.3), which gives:

$$p_{TH} = \frac{3R}{V_0^0} \sum_i \gamma_{i_0} a_i \left( \frac{\theta_i}{e^{\frac{\theta_i}{T}} - 1} + \frac{AT^2}{2} + \frac{BT^3}{3} \right) \tag{7.4}$$

From this point, the empirical description discovered by Grover *et al.* [14] is formulated along the reference isotherm as follows:

$$V_0 = V_0^0 - c \ln\left(\frac{K_{T_0}}{K_{T_0}^0}\right) \tag{7.5}$$

Following the same procedure detailed in Supplementary Note A, an isothermal description of the volume is obtained from Eq. (7.5). The effect of temperature is then accounted for based on Eq. (7.2) and (7.1), and the following description is finally obtained:

$$V = -cE_i^{-1}\left(E_i\left(-\frac{V_0^0}{c}\right) - \frac{1}{K_{T_0}^0}\exp\left(-\frac{V_0^0}{c}\right)(p - p_{TH} - p^0)\right) \tag{7.6}$$

where it is stressed out that the $c$ parameter has to be a constant, as otherwise the product $\alpha K_T$ would not be pressure independent.

It is emphasized here that the same general behavior could be obtained from either Eq. (7.6) or from the model proposed by Anand et al. [20] based on the isothermal formulation from Saxena [21] of the Grover empirical law. Indeed, although Anand et al. [20] considered that the product $\alpha K_T$ was temperature independent, Eq. (7.4) could very well be injected in the authors' framework to generalize it. Therefore, the main difference lies in the isothermal equation of state that was used to build each model. In the formulation provided by Saxena [21] and used by Anand et al. [20], it appears that the bulk modulus was considered to vary linearly with pressure. An advantage of this simplifying assumption is that an explicit expression of $K_T$ can be obtained as a function of $T$ and $p$. In this work, the formulation of the empirical law from Grover et al. [14] presented in Eq. (7.5) was preferred to build the model so that it could be more precise.

To investigate on the features of the present model, it was applied to β-Sn. All the parameters of Eq. (7.3) were taken from [24] in addition to $K_{T_0}^0$ from Eq. (7.6). The $c$ parameter was adjusted based on the bulk modulus data along the atmospheric pressure isobar [32,48–50] critically selected in [24], and a value of $2.1 \times 10^{-6}$ was obtained. A satisfying fit was reached as presented in Supplementary Note H. However, above 1 GPa, the volume and bulk modulus data available along the reference isotherm could not be satisfactorily reproduced, as also highlighted in Supplementary Note H. This difficulty to reproduce the data accurately is due to the fact that there is only a single degree of freedom, that is the parameter $c$, to account for the variations of $K_T$ along both isobars and isotherms. It is similar to what was discussed in Section 5. It is a limitation of the present approach and of the similar Anand-Saxena-Grover model.

Then, $\alpha$, $K_T$, and the product $\alpha K_T$ of β-Sn are calculated along different isobars and the results are presented in Fig. 6(a-c). To begin with, it can be seen from Fig. 6(a) that the bulk modulus decreases strongly with temperature. As a result, at low pressures, it quickly becomes negative. Brosh *et al.* [11] obtained similar results in their modeling of solid Al using a Mie-Grüneisen equation of state also based on the use of thermal pressure, and it was discussed by the authors that this behavior meant that the phase had become mechanically unstable from this point. A consequence of this trend is that, when $K_T$ reaches 0, the product $\alpha K_T$ follows, deviating from the input atmospheric pressure description as presented in Fig. 6(b). As a result, $\alpha$ diverges to infinitely high values approaching from this critical point as seen in Fig. 6(c). In the present case of β-Sn, this sharp increase of $\alpha$ starts from roughly 700 K at $10^5$ Pa, which is only slightly higher than the corresponding melting point of the phase of 505 K. In the case of CaO however, which was also investigated based on the description proposed in [24], at $10^5$ Pa this increase starts before the melting point of the compound of 3222 K [76]. As a result, the available high temperature thermal expansion data cannot be closely reproduced. Besides, it follows from Eq. (2.1) that this steep increase in the thermal expansion coefficient will lead to a dramatic decrease in the heat capacity with increasing pressure. While it can be argued that these extrapolations may not be physically meaningful if the phase had become mechanically unstable, this behavior occurs at low pressure at temperatures which are close to the melting point of the phases, and it appears as an important limitation in the context of establishing Gibbs energy databases.

It is interesting to note that, for the approaches based on the use of thermal pressure discussed in this section, the higher the pressure, the better the result. Indeed, it can be seen from Fig. 6 that for β-Sn, problematic features are obtained at low pressures, whereas reasonable extrapolations are achieved at 100 GPa. In the models investigated so far, it is rather at high pressure that problems would arise. Some insights on this unusual feature can be gained from Fig. 6(d), where $p - p^0 - p_{TH}$ is presented as a function of $T$ along different isobars. At $10^5$ Pa, this term is always negative. At the corresponding melting point of β-Sn, it reaches roughly -1.6 GPa. Yet, most isothermal equations of state, such as the one built based on the Grover empirical law in this section, were established for compression only. Therefore, it may be suggested that problematic results are obtained at low cold pressures and high temperatures because high tensile stresses are then input in an isothermal equation of state that is not valid in this domain. At very high cold pressures however, $p - p^0 - p_{TH}$ is positive, and reasonable results are thus obtained.

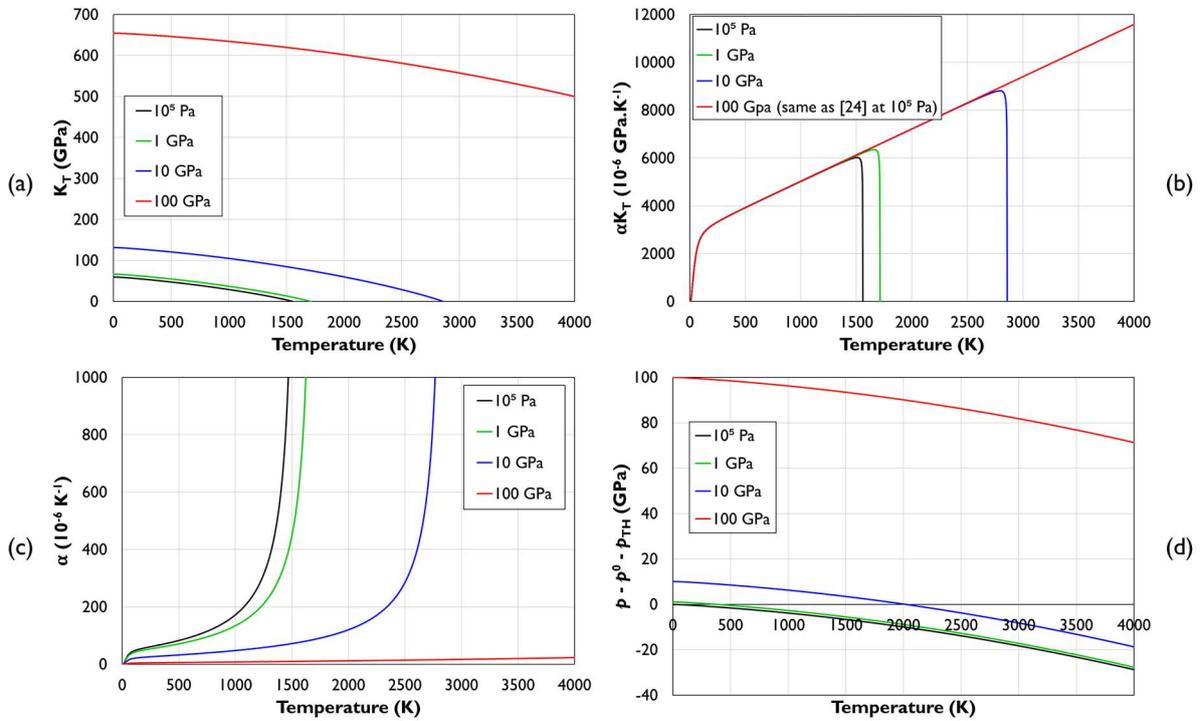

Fig. 6 – (a) Bulk modulus, (b) product $\alpha K_T$, and (c) thermal expansion coefficient of β-Sn calculated along different isobars using the approach based on the Grover empirical law and the use of thermal pressure presented in Section 7. (d) Overall pressure input in the model calculated from the difference between the cold pressure and the thermal pressure.

## 8. A new scheme based on the equivalence between pressure and temperature

For the models based on the use of thermal pressure discussed in Section 7, it was highlighted that problematic results were obtained when $p - p^0 - p_{TH}$ was significantly negative. Nevertheless, satisfying extrapolations were reached for high cold pressures. Therefore, in the present section, the same model as presented in Section 7 is adopted when $p - p^0 - p_{TH}$ is positive. However, when $p - p^0 - p_{TH}$ is negative, a new scheme also based on the equivalence between pressure and temperature when calculating the volume is proposed. A corollary of the approach leading to Eq. (7.2) is that the volume at a given high temperature and pressure is equivalent to a volume along the atmospheric pressure isobar, but at a lower temperature. Put into equation, it gives:

$$V(T,p) = V(T_{EQUIV}, p^0) \tag{8.1}$$

with $T_{EQUIV}$ the equivalent temperature defined when $p - p^0 - p_{TH}$ is negative as:

$$p^0 + p_{TH} - P = \int_{T_0}^{T_{EQUIV}} \alpha K_T dT \tag{8.2}$$

In this study, an explicit expression of $T_{EQUIV}$ as a function of $T$ and $p$ could not be obtained. Therefore, this term had to be computed numerically from the function $p_{TH} = f(T)$ as $T_{EQUIV} = f^{-1}(p - p^0 - p_{TH})$. This makes this model less computationally efficient than the other approaches previously discussed.

To summarize, when the difference between the cold pressure and the thermal pressure is negative (i.e., at rather low $p$ or high $T$, see Fig. 6(d)), the volume is calculated along the $p^0$ isobar based on Eq. (8.1) and (8.2). Because the product $\alpha K_T$ is considered to be pressure independent, all that is required in this range is $\alpha^0$, $K_T^{\ 0}$, and $V_0^0$, i.e., a description at atmospheric pressure of the volume and related properties. Then, when the difference between the cold pressure and the thermal pressure becomes positive (i.e., at rather high $p$ or low $T$, see Fig. 6(d)), the volume is calculated along the room temperature isotherm based on Eq. (7.6). An additional parameter $c$, which is a constant, is required in this range.

This scheme was applied to the modeling of β-Sn. The description of the volume and related properties at atmospheric pressure was taken from [24], and the thermal pressure was then calculated using Eq. (7.4) that is derived from the authors' model. Next, the $c$ parameter from Eq. (7.6) was adjusted using the bulk modulus [53–56] and molar volume [34,35,51,52] data available along the room temperature isotherm. As in Section 6, a value of 3.5x10$^{-6}$ was obtained for this parameter.

On this basis, the product $\alpha K_T$, $\alpha$, and $K_T$ of β-Sn were calculated along different isobars, and the results are presented in Fig. 7. Reasonable extrapolations of $\alpha$ and $K_T$ are obtained up to very high temperatures at both low and high pressures, as shown in Fig. 7(b-c). Plus, the available experimental data along both the atmospheric pressure isobar and the room temperature isotherm were fitted closely, as shown in Supplementary Note I. This satisfying fit was reached under the assumption that the product $\alpha K_T$ was pressure independent, and without using an extra $a$ parameter as in the Jacobs-Grover framework. However, at the specific conditions of $p$ and $T$ where $p - p^0 - p_{TH}$ equals 0, such as at 10 GPa and roughly 2000 K, a temperature derivative discontinuity is observed for $\alpha$ and $K_T$. This is due to the transition between the two submodels when the overall pressure goes from tensile to compressive. It may be argued that, in practice, this breakpoint in the temperature derivative of $\alpha$ and $K_T$ should not significantly impact the validity of the description.

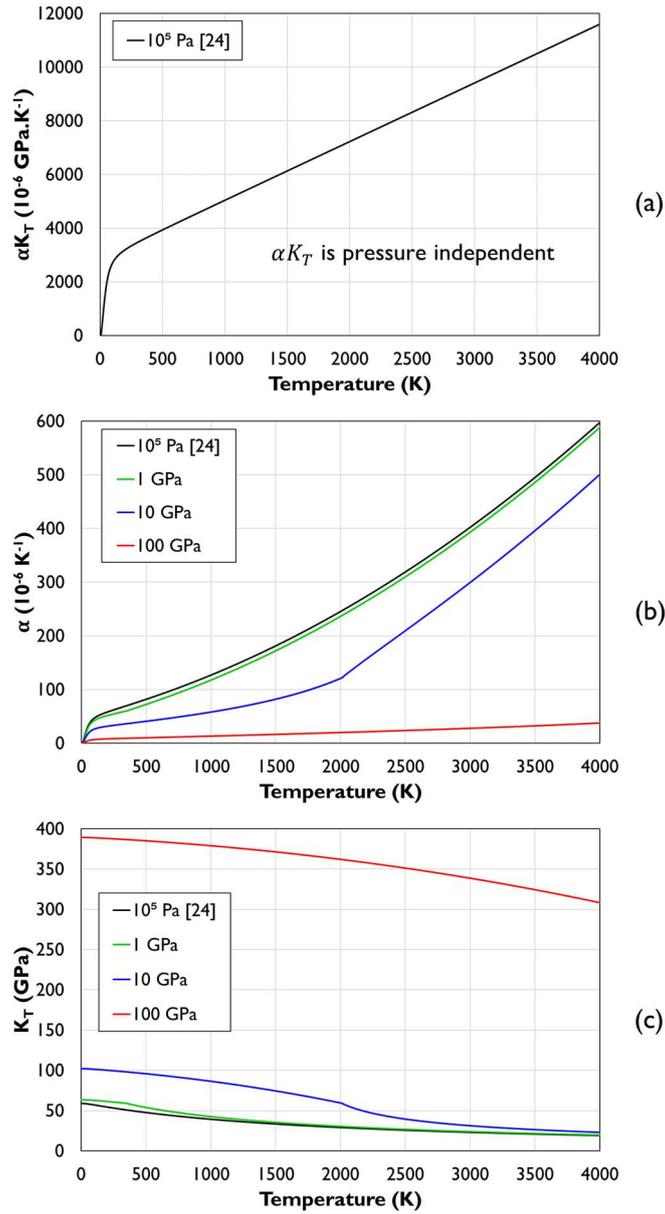

Fig. 7 – (a) Product $\alpha K_T$, (b) thermal expansion coefficient, and (c) bulk modulus of β-Sn calculated along different isobars using the new scheme presented in Section 8 that is based on the equivalence between temperature and pressure when calculating the volume.

In Fig. 8, the heat capacity of β-Sn is calculated numerically along different isobars based on the present description of the volume and on the atmospheric pressure thermodynamic description from [57]. A reasonable description is obtained up to temperatures and pressures of roughly twice the atmospheric pressure melting point and the standard bulk modulus of the phase, respectively. However, the heat capacity decreases too much with increasing pressure, and the model does not converge to the Dulong-Petit limit of 3R at high temperatures and pressures. Additional constraints in the model or in the optimization procedure would be required to improve the description. It is noted that very slightly negative values are obtained below 30 K.

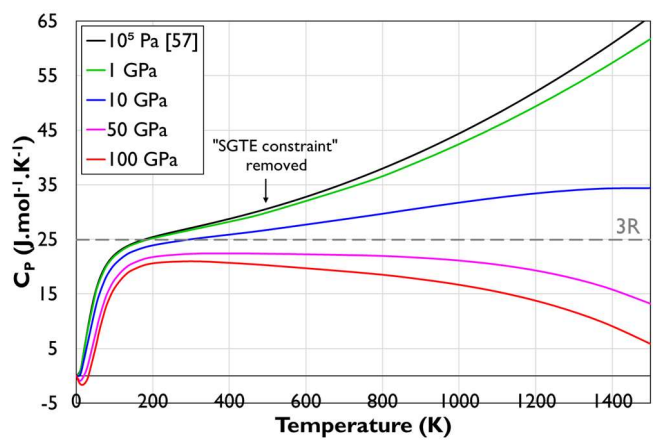

Fig. 8 – Heat capacity of β-Sn calculated numerically from the description of the volume presented in Section 8 and the atmospheric pressure thermodynamic description from [57] from which the "SGTE constraint" was removed.

## Conclusion

The current shortcomings of high-pressure Gibbs energy approaches are identified to be caused by an inconsistent treatment of the SGTE method of extrapolation, and an insufficient knowledge of the underlying models to describe the volume and their limitations. Two possible practical ways of solving the first problem were outlined in Section 2. Then, several models to describe the volume were analyzed.

First, it was demonstrated in Section 3 that the Lu-Grover model leads to an unphysical description of the thermal expansion at high pressure. This model is only applicable at relatively low pressures that should not exceed one fourth of the standard bulk modulus of the material. In the Joubert-Lu-Grover model discussed in Section 4, new parameters are added to improve the description. This revised model has been applied successfully to describe the volumetric properties of several metals up to high temperatures and pressures. Within this framework, the thermodynamic properties can be computed in two ways. The first is by performing a numerical integration of $V$ over $p$, which leads to an overestimated heat capacity at high pressure. This could possibly be improved by modifying the mathematical expression of the pressure cut-off terms introduced in the model. The second is by using an equation that leads to an approximate description of the heat capacity. This approach allows direct compatibility with the SGTE method of extrapolation. It appears promising to extend CALPHAD databases up to pressures of at least twice the standard bulk modulus of the phases, and temperatures higher than twice their atmospheric pressure melting point. Further studies, notably on materials other than metals, would be of interest.

Then, the Jacobs-Grover model was investigated in Sections 5-6, and improvements were proposed to make it more computationally efficient and to avoid an unphysical high-pressure feature. It was found that the Jacobs-Grover approach is only applicable within a limited temperature range, and that the higher the pressure, the lower the critical temperature from which problematic results are obtained. At pressures close to the standard bulk modulus of the material, the Jacobs-Grover model should not be applied at temperatures higher than its atmospheric pressure melting point.

Last, a model based on the use of thermal pressure to account for the temperature dependence of the volume was investigated in Section 7. This attempt leads to the same general behavior as the Anand-Saxena-Grover model. The data on the volumetric properties of β-Sn could not be satisfactorily reproduced by this approach. Besides, problematic results were obtained at relatively low cold pressures and high temperatures. To solve both these problems, a new scheme based on the equivalence between $T$ and $p$ was proposed in Section 8. This model led to a reasonable description of the volumetric properties of β-Sn for all temperatures and pressures, but it relies heavily on numerical calculations. Besides, additional constraints in the model or in the optimization procedure would be required to improve the description of the heat capacity. Therefore, this approach seems to have potential, but requires further refinement and validation.

## Acknowledgements

The authors gratefully acknowledge the French consortium in high temperature thermodynamics GDR CNRS n°3584 (TherMatHT), where constructive exchanges led to the present collaborative work. This research did not receive any specific grant from funding agencies in the public, commercial, or not-for-profit sectors.

# Data Statement

For the models investigated in Section 3 and 4, thermodynamic database files made for the Thermo-Calc [66] software are provided as supplementary materials. The atmospheric pressure data that were used to constrain some model parameters are available in [77]. Other materials (literature datasets, worksheets…) can be made available upon request.

# Declaration of competing interest

We declare no competing interests.

# Supplementary Notes for:

On models to describe the volume in the context of establishing high-pressure Gibbs energy databases

Guillaume Deffrennes[a,*], Jean-Marc Joubert[b], Benoit Oudot[a,*]

[a] CEA, DAM, VALDUC, F-21120 Is-sur-Tille, France

[b] Univ. Paris Est Creteil, CNRS, ICMPE, UMR 7182, 2 rue Henri Dunant, 94320 Thiais, France

* Corresponding authors:

Dr. Guillaume Deffrennes

Present postal address: National Institute for Materials Science, 1-1 Namiki, Tsukuba, Ibaraki 305-0044, Japan

e-mail : guillaume.deffrennes@gmail.com

Dr. Benoit Oudot

Postal address: CEA, DAM, VALDUC, F-21120 Is-sur-Tille, France

e-mail : benoit.oudot@cea.fr

**Supplementary Note A: Derivation of the equations presented in section 3 (Lu-Grover model)**

**How to obtain Eq. (3.1) of the manuscript:**

The empirical relationship discovered by Grover et al. [1] was formulated by Lu et al. [2] as follows:

$$V = V^0 - c \ln\left(\frac{K_T}{K_T^{\,0}}\right) \qquad (A.1)$$

with $c$ a material characteristic parameter that can be temperature dependent, and $K_T$ the isothermal bulk modulus defined as the inverse of the compressibility $\kappa_T$ as follows:

$$K_T = -V \left(\frac{\partial p}{\partial V}\right)_T \qquad (A.2)$$

By re-arranging Eq. (A.1), and by injecting Eq. (A.2) into the expression, the following differential equation is obtained:

$$\frac{\exp\left(-\frac{V^0}{c}\right)}{K_T^{\,0}} = -\frac{\exp\left(-\frac{V}{c}\right)}{V}\left(\frac{\partial V}{\partial p}\right)_T \qquad (A.3)$$

Finally, by integrating Eq. (A.3) and re-arranging the resulting expression, Eq. (3.1) from the manuscript is obtained:

$$V = -c\,Ei^{-1}\left(Ei\left(-\frac{V^0}{c}\right) - \frac{1}{K_T^{\,0}}\exp\left(-\frac{V^0}{c}\right)(p - p^0)\right) \qquad (3.1)$$

where $Ei$ is the exponential integral function, that can be calculated numerically from tabulations, and that is defined as:

$$Ei(x) = \int_{-\infty}^{x} \frac{e^t}{t} dt \tag{A.4}$$

It is noted that Eq. (3.1) differs from the original equation (9) from Lu *et al.* [2], because the authors actually used the $E_1$ function, that was noted $Ei$, and that is defined only for positive value of $x$ as $E_1(x) = -Ei(-x)$.

**How to obtain Eq. (3.3) of the manuscript:**

The variations of the product $\alpha K_T$ within the Lu-Grover framework can be obtained starting from:

$$\left(\frac{\partial(\alpha K_T)}{\partial p}\right)_T = \alpha \left(\frac{\partial K_T}{\partial p}\right)_T + K_T \left(\frac{\partial \alpha}{\partial p}\right)_T \tag{A.5}$$

Then, from the thermodynamic identity:

$$\left(\frac{\partial \alpha}{\partial p}\right)_T = \frac{1}{K_T^2} \left(\frac{\partial K_T}{\partial T}\right)_p \tag{A.6}$$

, one obtains the generally valid relationship:

$$\left(\frac{\partial(\alpha K_T)}{\partial p}\right)_T = \alpha \left(\frac{\partial K_T}{\partial p}\right)_T + \frac{1}{K_T}\left(\frac{\partial K_T}{\partial T}\right)_p \tag{A.7}$$

In the Lu-Grover equation of state, the pressure derivative of the bulk modulus can be obtained by derivation of Eq. (A.1) as follows:

$$\left(\frac{\partial V}{\partial p}\right)_T = -\frac{c}{K_T}\left(\frac{\partial K_T}{\partial p}\right)_T \tag{A.8}$$

Then, by injecting Eq. (A.2) into Eq. (A.8), one obtains:

$$\left(\frac{\partial K_T}{\partial p}\right)_T = \frac{V}{c} \tag{A.9}$$

The temperature derivative of the bulk modulus can be determined in a very similar manner by derivation of Eq. (A.1), and by injecting the definition of the thermal expansion coefficient into the resulting expression. One finally obtains:

$$\left(\frac{\partial K_T}{\partial T}\right)_p = \frac{K_T}{c}(V^0\alpha^0 - V\alpha) + \frac{K_T}{K_T^0}\left(\frac{\partial K_T^0}{\partial T}\right)_p - \frac{K_T}{c}\ln\left(\frac{K_T}{K_T^0}\right)\left(\frac{\partial c}{\partial T}\right)_p \tag{A.10}$$

By injecting Eq. (A.9) and (A.10) into Eq. (A.7), Eq. (3.3) from the manuscript is obtained:

$$\left(\frac{\partial \alpha K_T}{\partial p}\right)_T = \frac{V^0\alpha^0}{c} + \frac{1}{K_T^0}\left(\frac{\partial K_T^0}{\partial T}\right)_p - \frac{1}{c}\ln\left(\frac{K_T}{K_T^0}\right)\left(\frac{\partial c}{\partial T}\right)_p \tag{3.3}$$

**Supplementary Note B: Description at atmospheric pressure of the volume, thermal expansion and heat capacity of β-Sn accepted in this work compared with experiments**

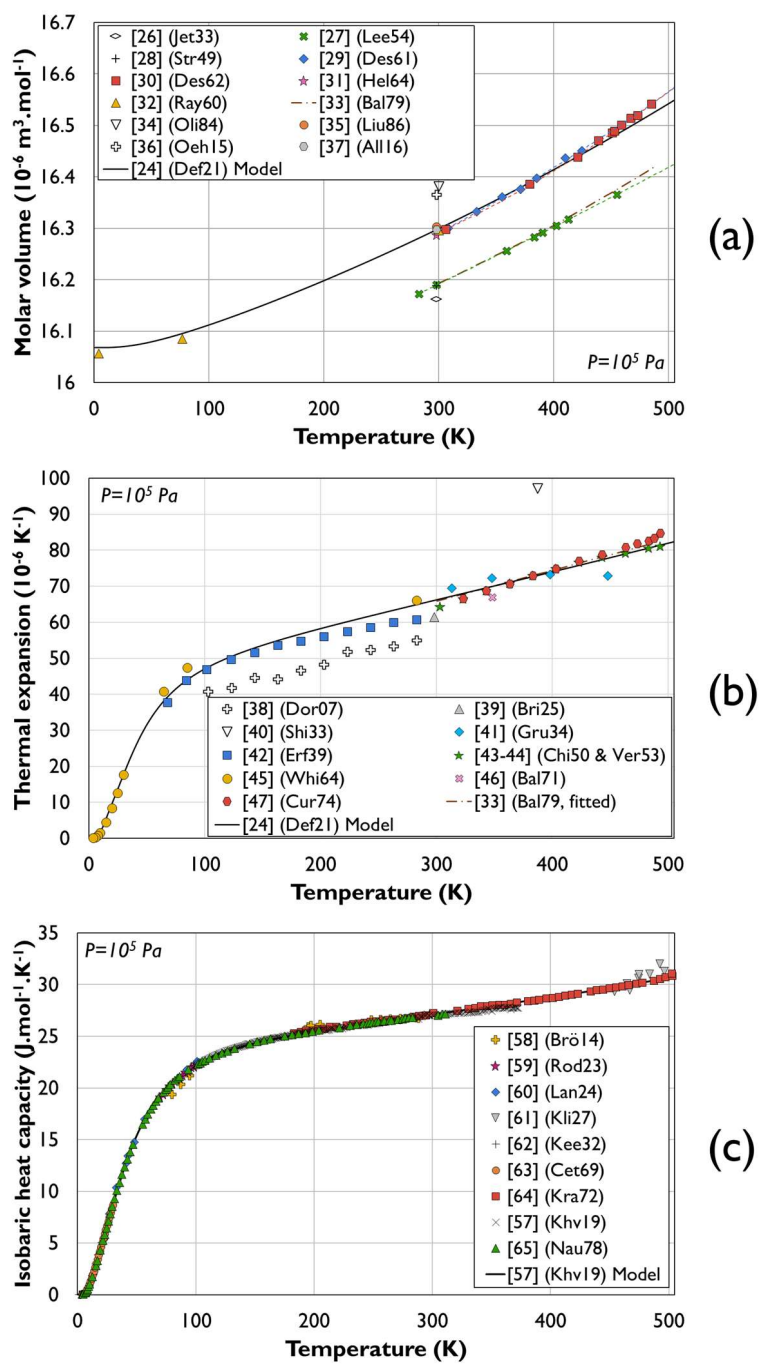

**Fig. S1** – Atmospheric pressure description of (a) the volume, (b) the thermal expansion, and (c) the heat capacity of β-Sn that is used in the various models investigated in this work compared with experimental data. The description of the heat capacity is further supported by heat content data as shown in [4].

**Supplementary Note C: Agreement between calculations and experimental data for β-Sn when the Lu-Grover model presented in Section 3 is used**

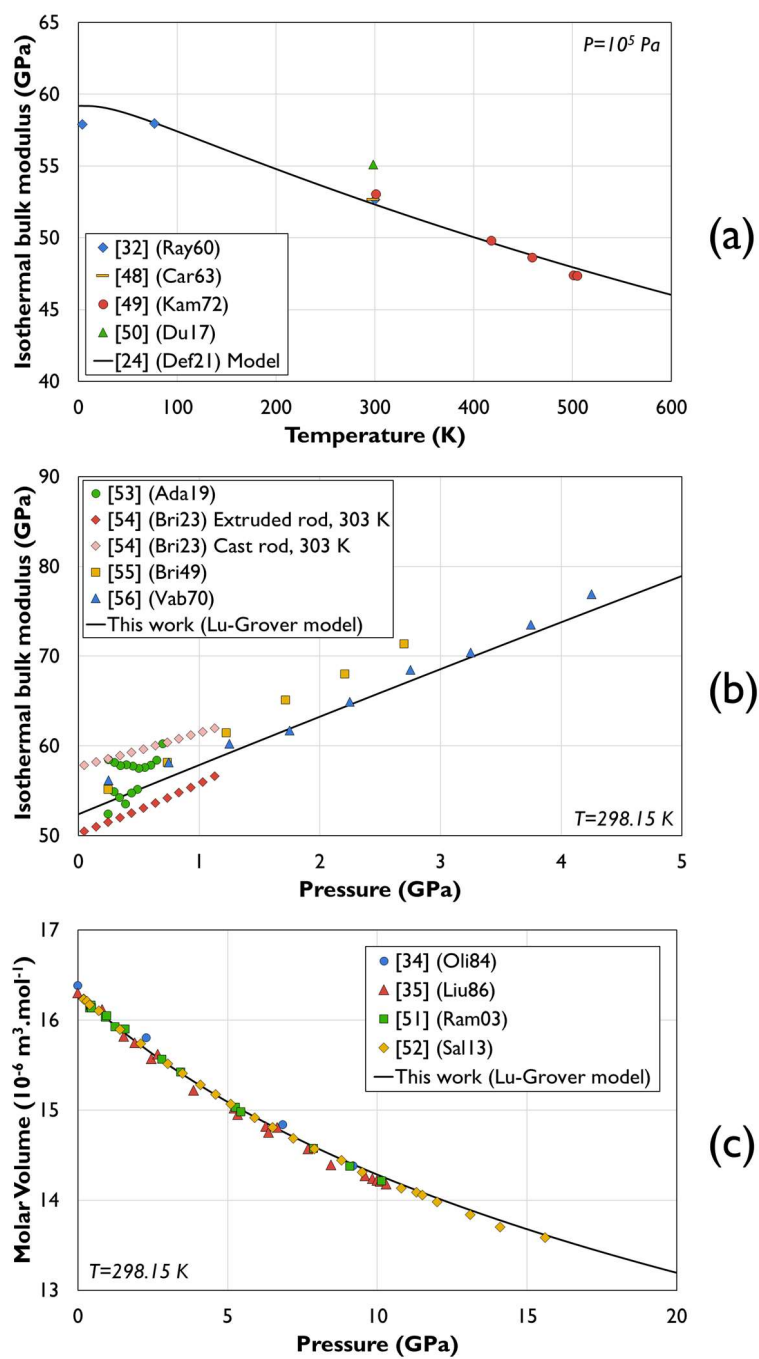

**Fig. S2** – Bulk modulus along (a) the $10^5$ Pa isobar and (b) the 298.15 K isotherm and (c) molar volume along the 298.15 K isotherm as calculated using the Lu-Grover model compared with experimental data

**Supplementary Note D: The Joubert-Lu-Grover approach (Section 4) adapted to the 3rd generation model for the thermal expansion and bulk modulus from [3]**

First of all, following Eq. (4.2) of the manuscript, a cut-off parameter was applied to the parameter governing the temperature dependence of the bulk modulus as follows:

$$K_T^0 = \frac{1}{\chi_{T_0} + C(p) \sum_i \frac{a_i}{\exp\left(\frac{\theta_i}{T}\right) - 1}} \quad (C.1)$$

$$C(p) = C \exp\left(-\frac{p}{p_{CUT}}\right) \quad (C.2)$$

with $\chi_{T_0}$ the compressibility at the reference temperature, $C$ a material-dependent parameter that does not vary with $T$ and should not be confused with the Lu-Grover parameter $c$, $\theta_i$ the Einstein temperature associated with the i$^{\text{th}}$ Einstein mode of vibration, and $a_i$ the corresponding pre-factor.

Then, in [3], the thermal expansion coefficient is expressed from the description of the bulk modulus and of the isochoric heat capacity. The first cutoff parameter noted $p_{CUT}'$ in Eq. (4.1) and applied to the constant parameter in the 2$^{\text{nd}}$ generation description was transposed to the harmonic contribution to the heat capacity, which is also constant above the Einstein temperature. The second cutoff parameter noted $p_{CUT}$ in Eq. (4.1) was applied to the anharmonic and electronic contributions to the heat capacity, which gives the increase in the thermal expansion coefficient at high temperature. Finally, the following expression was obtained:

$$\alpha(T) = \frac{3R}{V_0} \sum_i \gamma_{i_0} a_i \left( \left(\frac{\theta_i}{T}\right)^2 \frac{e^{\frac{\theta_i}{T}}}{\left(e^{\frac{\theta_i}{T}} - 1\right)^2} \exp\left(-\frac{p}{p_{CUT}}\right) \left( \chi_{T_0} + \frac{C(p)}{e^{\frac{\theta_i}{T}} - 1} \right) \right.$$

$$\left. + \left( (AT + BT^2) \exp\left(-\frac{p}{p_{CUT}}\right) \left( \chi_{T_0} + C(p) \left(\frac{T}{\theta_i} - \frac{1}{2}\right) \right) \right) \right) \quad \text{(C.3)}$$

with $R$ the gas constant, $V_0$ the molar volume at the reference temperature, $\gamma_{i_0}$ the Grüneisen parameter associated with the i[th] Einstein mode of vibration, and $A$ and $B$ phenomenological parameters to account for anharmonic and electronic contributions to the heat capacity.

**Supplementary Note E: Agreement between calculations and experimental data for β-Sn when the Joubert-Lu-Grover model presented in Section 4 is used**

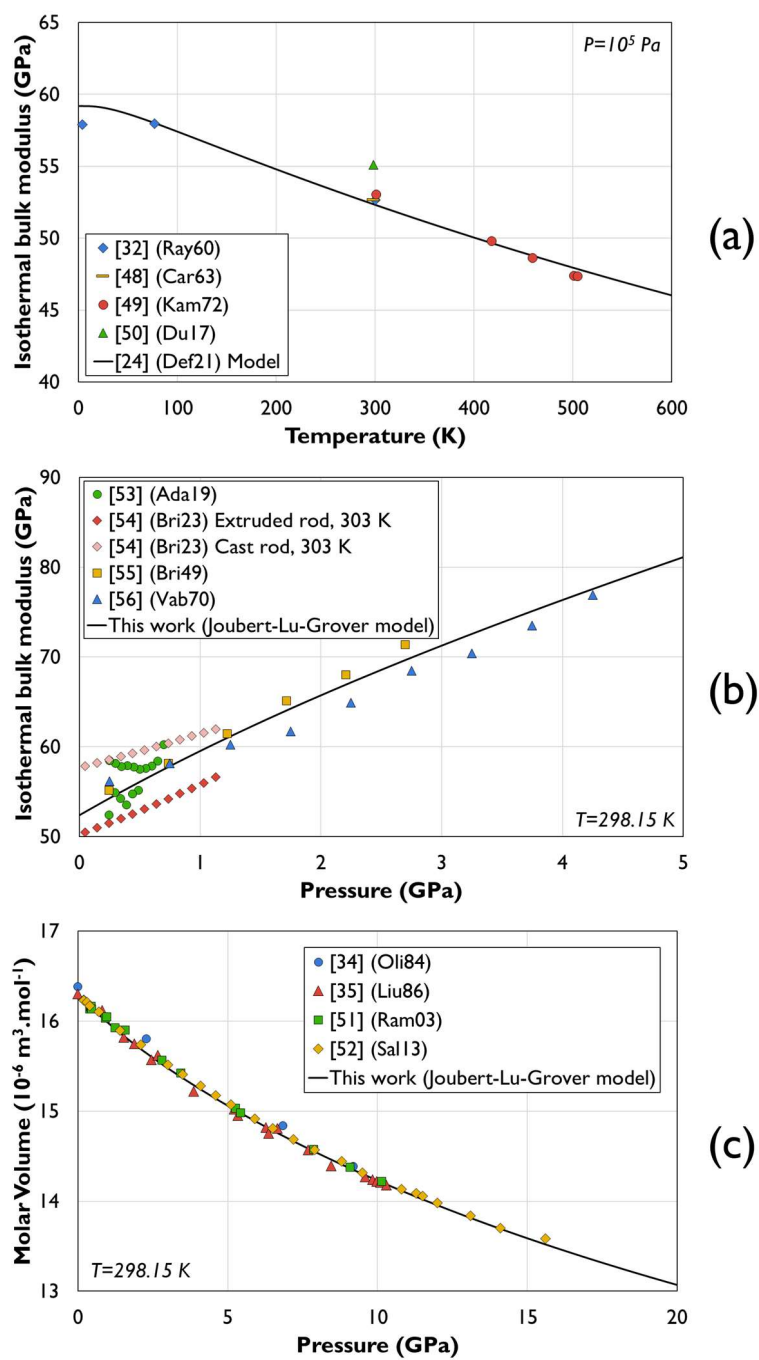

**Fig. S3** – Bulk modulus along (a) the $10^5$ Pa isobar and (b) the 298.15 K isotherm and (c) molar volume along the 298.15 K isotherm as calculated using the Joubert-Lu-Grover model compared with experimental data

**Supplementary Note F: Derivation of the equations presented in Section 5 (Jacobs-Grover model)**

**How to obtain Eq. (5.5) of the manuscript:**

Following the same approach that previously led to Eq. (A.9), it can be shown starting from Eq. (5.2) of the manuscript that the expression of $(\partial K_T/\partial p)_T$ is unchanged from the framework of the Lu-Grover equation of state.

Then, the temperature derivative of the bulk modulus is obtained by derivation of Eq. (5.2), and by injecting the definition of the thermal expansion coefficient into the resulting expression, the following equation is obtained:

$$\left(\frac{\partial K_T}{\partial T}\right)_p = \frac{K_T}{c}(a - V\alpha) \tag{E.1}$$

Finally, by injecting Eq. (A.9) and (E.1) into Eq. (A.7), Eq. (5.5) of the manuscript is obtained.

**How to obtain Eq. (5.6) of the manuscript:**

Eq. (5.6) of the manuscript can be either obtained by derivation of Eq. (5.2), or by injecting Eq. (A.9) and (E.1) into the thermodynamic identity:

$$\left(\frac{\partial K_T}{\partial T}\right)_V = \left(\frac{\partial K_T}{\partial T}\right)_p + \alpha K_T \left(\frac{\partial K_T}{\partial p}\right)_T \tag{E.2}$$

**Supplementary Note G: Agreement between calculations and experimental data for β-Sn when the revised Jacobs-Grover model presented in Section 6 is used**

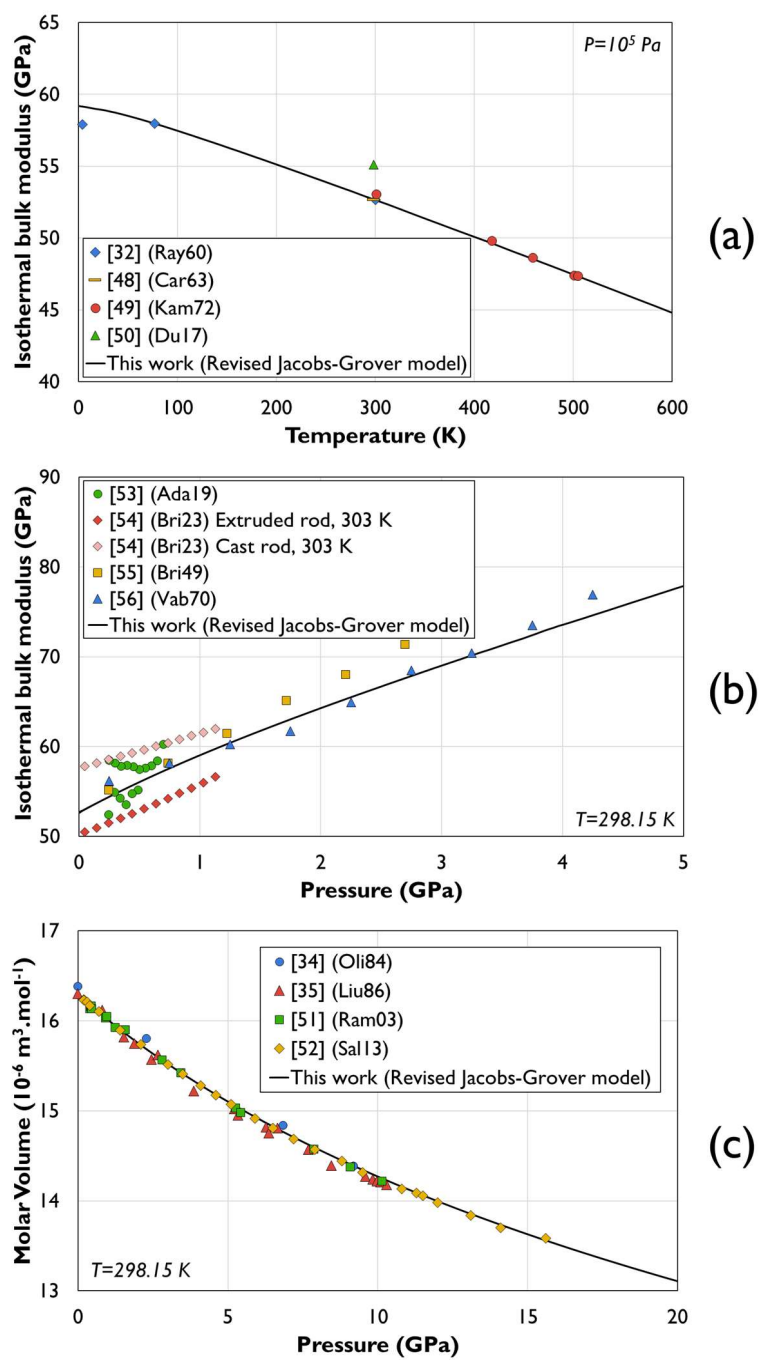

**Fig. S4** – Bulk modulus along (a) the $10^5$ Pa isobar and (b) the 298.15 K isotherm and (c) molar volume along the 298.15 K isotherm as calculated using the revised Jacobs-Grover model compared with experimental data

**Supplementary Note H: Agreement between calculations and experimental data for β-Sn when the model based on thermal pressure presented in Section 7 is used**

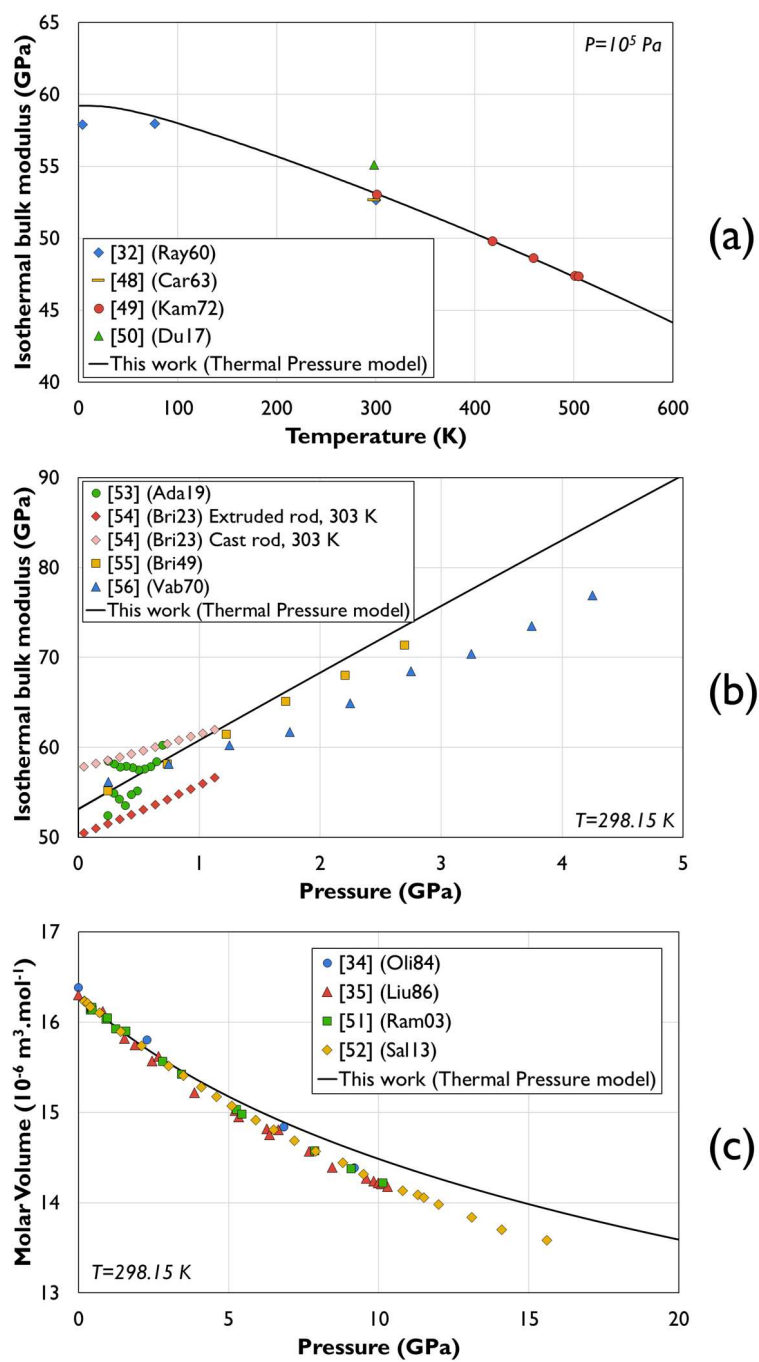

**Fig. S5** – Bulk modulus along (a) the $10^5$ Pa isobar and (b) the 298.15 K isotherm and (c) molar volume along the 298.15 K isotherm as calculated using the model based on thermal pressure presented in Section 7 compared with experimental data

**Supplementary Note I: Agreement between calculations and experimental data for β-Sn when the new scheme presented in Section 8 is used**

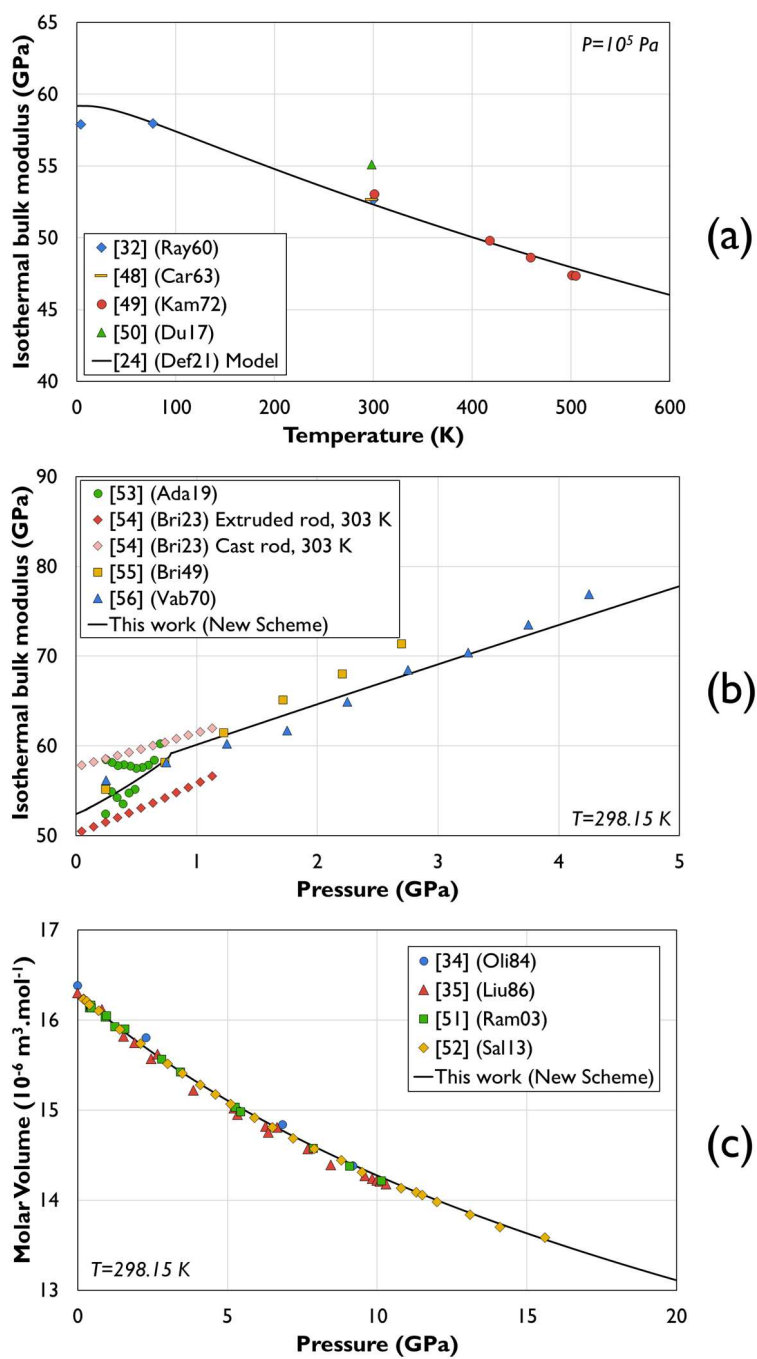

**Fig. S6** – Bulk modulus along (a) the $10^5$ Pa isobar and (b) the 298.15 K isotherm and (c) molar volume along the 298.15 K isotherm as calculated using the new scheme presented in section 8 compared with experimental data. The pressure derivative discontinuity observed in (b) for $K_T$ is discussed in more details in the manuscript (Section 8).